\documentclass[prb,nopacs,preprint,preprintnumbers,amsmath,amssymb,showpacs,showkeys]{revtex4-1}
\usepackage{graphicx}
\usepackage{amssymb, amsmath, amsbsy}
\usepackage{color}
\usepackage{bm}

\begin{document}
\preprint{Hrms/ArtGrAZ.dft/dftv1-1.tex}


\title{
Density functional theory analysis of flexural modes, elastic constants, and corrugations
in strained graphene.
}

\author{P.L. de Andres}
\author{F. Guinea}
\author{M.I. Katsnelson}

\affiliation{
Donostia International Physics Center,
Paseo Manuel de Lardizabal 4, 20018 Donostia, Spain.
}

\affiliation{
Instituto de Ciencia de Materiales de Madrid (CSIC),
Cantoblanco, 28049 Madrid, Spain.
}

\affiliation{
Radboud University Nijmegen, Institute for Molecules and Materials,
Heyendaalseweg 135, 6525AJ NIjmegen, The Netherlands.
}

\date{\today}

\begin{abstract}
{\it Ab initio} density functional theory has been used to analyze
flexural modes, elastic constants, and atomic corrugations on single and
bi-layer graphene.
Frequencies of flexural modes are sensitive to
compressive stress; its variation under stress can be
related to the anomalous thermal expansion via a simple model
based in classical Elasticity Theory.\cite{AGK12}
Under compression, flexural modes are responsible for a long wavelength
rippling with a large amplitude and a marked anharmonic behavior.
This is compared with corrugations
created by thermal fluctuations and the adsorption of a light impurity (hydrogen).
Typical values for the later are in the sub-Angstrom regime, while
maximum corrugations associated to bending modes
quickly increase
up to a few Angstroms under a compressive stress,
due to the intrinsic instability of flexural modes
\end{abstract}






\maketitle

\section{Introduction.}


Physical properties of single and multilayered graphene are remarkable and
open interesting avenues for applications in the future.\cite{NGPNG09}
However, to realize its potential it is important to fully understand
the elastic properties of
graphene layers,\cite{lee08} the effect of external stresses on
electronic properties\cite{andres08AA,AAG11},
or conversely the effect of doping on mechanical properties,\cite{duan12}
and the appearance of
ripplings,\cite{FLK07,SGG11} corrugations,
and other defects expected under actual conditions
of cleanness, temperature, and growth.
The understanding of any departure from the perfect 2D configuration
(e.g., rippling and corrugation) of very thin graphene layers,
either in free suspension or deposited on a template substrate,
is central for manufacturing processes.\cite{lau09,morgenstern10}
At T= 0 K, in the absence of defects,
the carbon bond on the graphene layer is well understood in terms of
the formation of three in-plane localized strong sp$^{2}$ bonds, and
a fourth delocalized, out-of-plane, $\pi$-like bond.\cite{NGPNG09,K12}
The optimum geometrical configuration is achieved by
a honeycomb lattice formed by two equivalent sublattices
displaying P6/mmm symmetry.
The corresponding electronic structure shows bands
dispersing linearly around the Fermi energy that are
responsible for the fast and efficient transport of carriers.
Both experimentally and theoretically,\cite{lee08,mounet05,ZKF09} it is
shown that this kind of arrangement results in a material with the largest
in-plane elastic constants known yet.
Therefore the 2D perfect flat layer makes the most stable configuration since
deviations from a common plane requires a significant amount of energy.
Any departure from such a scenario affects greatly the atomic scale
properties of the layer and must be understood in order to efficiently exploit
graphene's properties.
Different reasons, however, might be invoked for a two-dimensional graphene layer
to adopt a
certain corrugation at different scales. First, at a non-zero temperature
there is a thermodynamic argument that implies the impossibility
for a perfect 2D layer to exist
in 3D.\cite{landauSP1,NP87,AL88,doussal92,FLK07,K12}
Second, defects like adsorbed impurities, vacancies, etc,
create local corrugations at the atomic scale\cite{BKL08} that propagate via the
elastic properties of the lattice
originating long-range correlations.
Finally, external applied stresses related to conditions on the boundary
make graphene to bend and to corrugate;
an interesting point to study since the growth of graphene layers on different
supporting substrates implies mismatches that introduce
all kind of stresses that have been observed to originate
a highly complex and corrugated landscape.\cite{morgenstern10,KCNG11}
Research on these issues require
atomically resolved techniques;
theoretical simulations
performed with {\it ab-initio} Density Functional Theory (DFT)\cite{DFT1,DFT2}
give an accurate picture of the interatomic interactions and
can provide valuable information about geometry, vibrations,
electronic effects, etc, down to the atomic level.
Indeed, {\it ab-initio} DFT has been successfully applied for many years now
to describe the different allotropes of
carbon, like graphite,
diamond,
fullerenes,
organic molecules,
nanotubes, etc (e.g. see refs. \cite{portal99,mounet05,cohen06}).
Here,
we focus on the role of external stresses on the flexural modes
ultimately responsible for the intrinsic thermodynamic instability of
these layers. The frequencies of these modes are very sensitive
to external compressive stresses and can be related, in turn, to the
Gr\"uneisen coefficients and the thermal expansion coefficient of graphene.\cite{AGK12}
The purpose of this paper is to use
state-of-the-art {\it ab-initio} DFT
to compare
atomic corrugations on graphene layers related to three different
origins:
(i) the presence of small adsorbed impurities (hydrogen),
(ii) thermal vibrational
amplitudes (T=300 K), and (iii) the long wavelength corrugation characteristic
of flexural modes.
We analyze the consequences of an isotropic compressive stress on these
corrugations, the elastic constants and the phonons.

\section{Ab-initio Density Functional Atomistic Model.}

\begin{figure}
\includegraphics[clip,width=0.99\columnwidth]{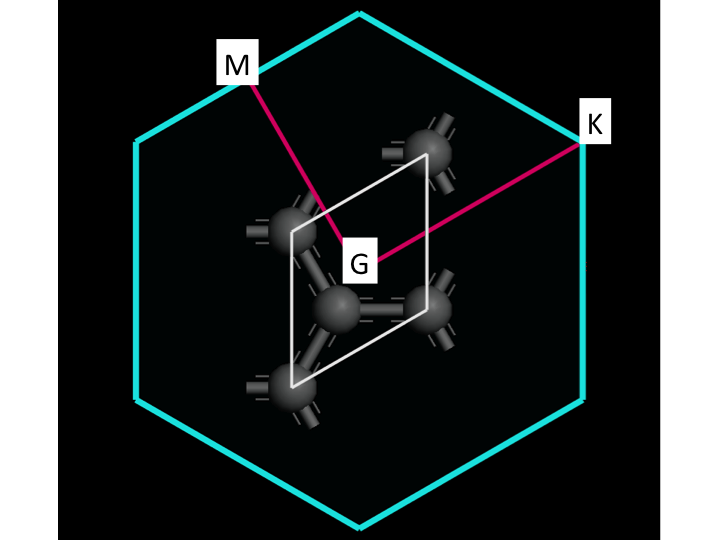}
\caption{
(Color online)
Graphene rhombohedral $1 \times 1 $ unit cell ($\gamma=60$), and
corresponding Brillouin Zone (special points are labelled
according to the phonon calculation in Fig. \ref{fgr:PH}).
\label{fgr:UCBZ}
}
\end{figure}

Since we are interested in elastic and vibrational properties we use
accurate norm-conserving pseudopotentials
to describe carbon valence band electrons
(2s$^{2}$ 2p$^{2}$).\cite{lee96,pseudo}
Electronic wavefunctions have been expanded in a plane-wave basis set
up to an energy cutoff of $750$\,eV, and a
$12 \times 12\times 2$ Monkhorst-Pack mesh
has been used to sample wavefunctions inside the first Brillouin zone.
Electronic bands were obtained using a smearing width of
$\eta=0. 1$\,eV.
The choice of the exchange and correlation (XC) potential opens up
various possibilities.
We favor the local density approximation (LDA)\cite{LDA} because of its
simplicity, excellent structural results for a wide range of
carbon allotropes relevant for this work, and in particular because,
even if somehow fortuitous, it predicts a reasonable value for
the layer separation in graphite.
LDA presents a solid record reproducing many key features for a
wide variety of carbonaceous materials, and it has been a standard
choice in the literature for many authors.
Unfortunately, the use of LDA cannot be justified "a priori" on physical grounds by
merely referring to the behavior of the next term on a gradients expansion (GGA);
different flavors of GGA yield worse agreement with experiments 
(e.g. for the distance between planes in graphite) showing that
the converge on such an expansion is not a trivial one. 
Therefore, we are forced to justify the use of LDA on the "a posteriori"
reasonable structural results. Its simpler formulation should also be taken
as an advantage, since makes a better defined and more unique formulation.
Finally, it is known that LDA may not be too accurate for absolute
values of total energies, but it affords a good description of relative
values (e.g., barriers), and it yields the correct dependence with
stress after a single common offset correction.\cite{andres08}
A periodic supercell with a vacuum gap in the perpendicular direction
($30$ {\AA}) and several $n \times m$ in-plane periodicities ($n,m= 1 - 8$)
depending on the requirements of each model have been used.
Geometrical parameters have been optimized so residual maximum
forces and stresses have been kept below
$F_{max} \le 0.0001$ eV/{\AA}, and $S_{max} \le 0.0002$ GPa
for the calculation of phonons and elastic constants,
yielding a lattice parameter for the rhombohedral unit cell on
a single graphene layer
of $a=b=2.447$ {\AA} (Fig. \ref{fgr:UCBZ}),
and $c=6.65$ {\AA} for graphite (Bernal stacking).
Calculations were performed with the CASTEP
code, as implemented in
Materials Studio.\cite{Clark05, accelrys}

The elastic tensor has been obtained by applying a series of strains on
the $1 \times 1$ unit cell (maximum amplitude $0.003$ {\AA}),
and by computing the associated stress tensors.\cite{strainstress}
Phonons have been computed using linear response theory.\cite{phononsL}
As a global benchmark to the elements included in our
calculations (pseudopotential, exchange and correlation model,
different convergence thresholds, etc), we compare in
Table I 
our theoretical results for the five
independent components of the elastic tensor for graphite to
experimental values measured on pyrolitic graphite
by ultrasonic resonance,\cite{cijgrPyr}
and on single crystals by inelastic
x-ray scattering.\cite{cijgrIXR}
A comparison to a recent independent theoretical determination
has also been included.\cite{SKF11}
The agreement is quite reasonable, except perhaps for $c_{13}$ that
cannot be described properly with a local choice for
exchange and correlation.
We notice that the basal in-plane theoretical Poisson ratio, $\nu_{xy}=\nu_{yx}=
\frac{c_{12}}{c_{11}} = 0.19$,
and the Young modulus,
$Y_{x}=Y_{y}=1$ (TPa),
are in good agreement with experimental values, $0.165$ and $1.1$ TPa
respectively;\cite{cijgrPyr} as expected these are very similar
for a graphene monolayer ($0.18$ for theory to be compared with 
a quoted experimental value $of 0.17$\cite{lee08}).

Values for the 2D elastic constants of a single graphene layer (I),\cite{gr2L1}
the bilayer with the usual AB Bernal stacking,
and a bilayer with direct AA stacking
(distance between layers $3.5$ {\AA})
display little variation in the main elements of the elastic
tensor (Table II, for reference 2D tensions for graphite
are quoted).
The main departure has been found
for the AA bilayer
in the {\it metastable} configuration
(distance between layers $1.5$ {\AA});\cite{andres08AA}
the reduction of c$_{11}$ is related to the gradual transformation
of the in-plane C-C bond order from sp$^{2}$ towards sp$^{3}$,
since the 2D unit cell needs to be expanded to make the structure
metastable under its own pressure.
The increase in c$_{66}$
implies a larger tendency against shear due to substitution of the
weak van der Waals-like interaction between both layers by stronger sp$^{3}$-like
chemical bonds.

For the two-dimensional graphene layer, due to its 6-fold symmetry, we
only need to determine two independent values in the elastic tensor;
these can be related to the two Lame coefficients needed to characterize
an isotropic system;
$\lambda=c_{12}$ and
$\mu=c_{66}$.
An independent determination of
$c_{11}$, that should be equal to $c_{12} + 2 c_{66}$,
holds fairly well in all the cases.
Therefore, simple analytical models based in the theory of
elasticity that only need values for these two parameters
make a good approximation and can be
thoroughly analyzed.\cite{landauEl,AGK12}
Under a moderate compressive stress (e.g. $\epsilon=-0.04$)
the main change happens in the value of
$c_{12}$ that is reduced by $\approx 2$.

Fig. \ref{fgr:PH} gives the phonons for graphene along
two main directions in the irreducible part of the
Brillouin Zone: G-M and G-K (Fig. \ref{fgr:UCBZ}).
Since there are two atoms in the unit cell,
we find three acoustic (ZA, TA and LA)
and three optical modes (ZO, TO and LO).
The optimized structure shows
a computed vibrational spectra that agree well with
previous calculations in the literature,\cite{saito} as can be seen by
comparing frequencies for the modes at selected high-symmetry points
(Table III).
For $\epsilon=0$ (no external stress, black circles),
all the frequencies are positive and behave as expected;
it is interesting to observe how the lowest acoustic mode (ZA)
disperses quadratically, a characteristic feature for a 2D system
(e.g., see Eq. (3) in ref.~\cite{AGK12}).

\subsection{Instability of flexural modes under compression.}
\begin{figure}
\includegraphics[clip,width=0.99\columnwidth]{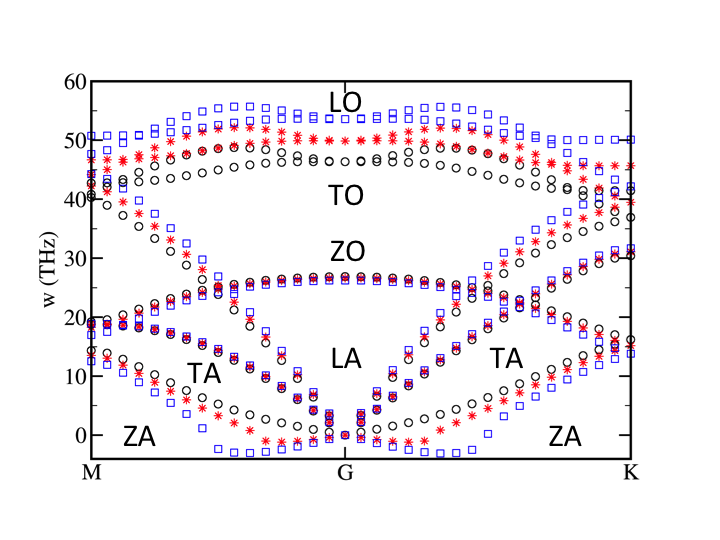}
\caption{
(Color online)
Phonons for the free-standing graphene layer (THz).
Path: $ M (0,\frac{1}{2},0) \rightarrow G (0,0,0)
\rightarrow K (\frac{1}{3},\frac{2}{3},0)$.
Black, circles: $\epsilon=0$.
Red, stars: $\epsilon=-0.02$.
Blue, squares: $\epsilon=-0.04$.}
\label{fgr:PH}
\end{figure}

The intrinsic instability of the layer against compressive
strain is demonstrated by the appearance of negative frequencies
in a small region around the G point;
the ZA phonon
softens and goes to negative values.\cite{liu07,zhang11}
It is worth noticing that because the extreme asymmetry
between compression and tension for strain induced instabilities on
graphene layers we have only considered compressive strains.\cite{zhang11}
Other failure mechanisms for graphene under tension have been
studied elsewhere.\cite{marianetti10}
For some critical wavevector, $q=q_{c}$, the mode becomes again positive,
and the value of $q_{c}$ increases with the amount of stress.
The value of $q_{c}$  defines a characteristic wavelength,
$\lambda_{c}=\frac{2\pi}{ q_{c} }$,
with an amplitude that depends directly on the strain.
This lowering of the frequency of the ZA modes under compression is
consistent with the negative Gr\"uneisen coefficients reported in the literature.\cite{mounet05,KF11,Petal11,AGK12}
A fit to the results in Fig.~\ref{fgr:PH} gives
$\omega_{\vec{\bf q}}^2 ( {\rm THz^2} ) \approx - 25 \left| \vec{\bf q} \right|^2 + 112 \left| \vec{\bf q} \right|^4$ for $\epsilon = 0.02$
and $\left| \vec{\bf q} \right|$ in \AA$^{-1}$ (valid near the center of the Brillouin Zone).
For a -2\% strain the corresponding 2D stress tensor is
$\sigma=\sigma_{xx}=\sigma_{yy}= 0.6$ eV/{\AA}$^{2}$,
$q_c \approx 0.35$ \AA$^{-1}$,
and $\lambda_{c} \approx 12$ {\AA}.
Other relevant feature under stress observed in Fig. \ref{fgr:PH}
is the hardening of LO and TO modes by $\approx 20$\% for $\epsilon=-0.04$.
These modes correspond to displacements
inside the plane of the layer,
play an important role in the dissipation of energy
after the absorption of electromagnetic radiation,\cite{Setal11,Getal11}
and are strictly degenerate at the G point.

Note, finally, that the softening of the ZA and ZO modes under strain, shown in Fig. \ref{fgr:PH} will enhance the intrinsic spin-orbit coupling of graphene.\cite{OFNG12}

\section{Corrugations under stress.}

\subsection{Flexural modes}
\begin{figure}
\includegraphics[clip,width=0.99\columnwidth]{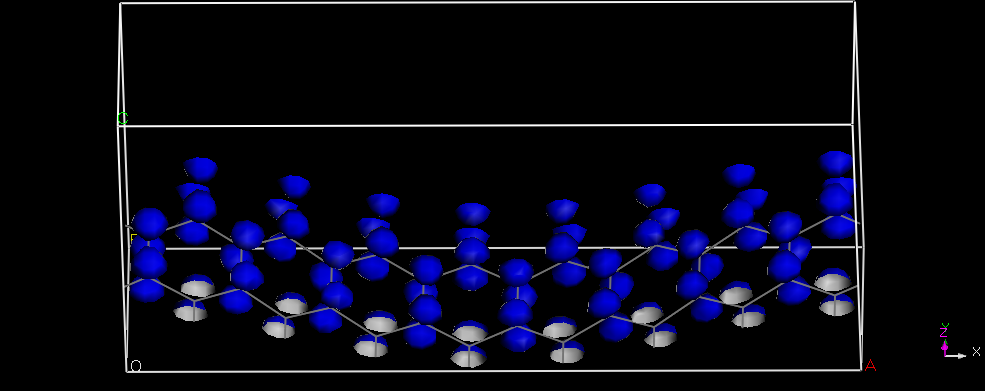}
\caption{
(Color online)
Long-wavelength deformation related to the AZ mode
computed on a
$8 \times 2$ rectangular unit cell ($\epsilon=-0.03$).
 Average and maximum corrugations are, $\overline{c}=0.66$,
and $c_{M}= 2.03$ {\AA},
respectively.
The density of states at $E_{F}+1$ V over the C atoms is
shown (STM image in the Tersoff-Haman approximation).
}
\label{fgr:STM}
\end{figure}

We address the question of deformations related to the intrinsic
instability brought by the flexural AZ mode.
Under compressive strain, the
negative region of frequencies near the G point
indicate the existence of an energetically favorable deformation for the system.
Therefore,
we essay a cosine-like perturbation with a wavelength corresponding to $q_{c}$
on a rectangular $8 \times 2$ supercell
under a range of strains from -2\% to -5\%.
These strains are large enough for the wavelength determined by the $8 \times 2$ supercell. 
It has been previously shown by Zhang and Liu\cite{zhang11}
that wavelengths of periodic undulations scale inversely with the square root of strain, 
setting a minimum value for a given wavelength that it is $< 1\%$ for the $8 \times 2$, 
therefore compatible with the values we use here. 
For $\epsilon=$-3\%, the soft mode displayed in
Fig. \ref{fgr:STM} allows the system to gain $-26$ meV with
a maximum amplitude of
$c_{M}= {\rm max} (z_{i}) - {\rm min} (z_{j}) = 2.03$ {\AA},
and an averaged corrugation of
$\overline{c} = \frac{1}{N} \sum_{i=1,N} \mid \overline{z} - z_{i} \mid = 0.66$ {\AA}.
Larger stresses drives the non-linear behavior of the system,
for $\epsilon=$-5\% we find an energy gain of $-3059$ meV
and $c_{M} =2.81$, $\overline{c} =0.92$ {\AA}.
It is interesting to notice that the density of states at the Fermi
level (e.g., the Tersoff-Haman STM image)
for such a long-range rippling of graphene is symmetric with respect to
the two sublattices, as can be seen in
Fig. \ref{fgr:STM}, where the density of states has been drawn on
each C atom for $V= 1$ V.
Corrugations of this kind are expected to evolve smoothly into
long wavelength deformations in larger unit cells,
unlike the case for thermal or defect induced corrugations we study below.

\subsection{Thermal fluctuations and light impurity adsorption.}

\begin{figure}
\includegraphics[clip,width=0.99\columnwidth]{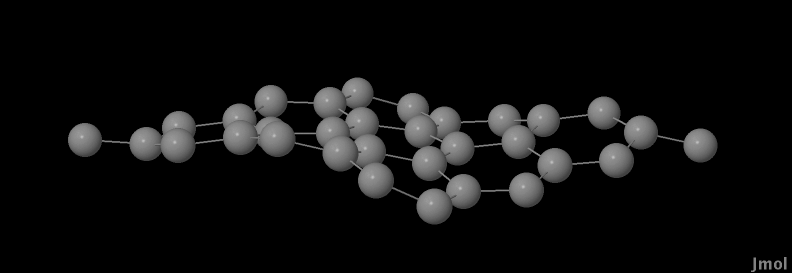}
\includegraphics[clip,width=0.99\columnwidth]{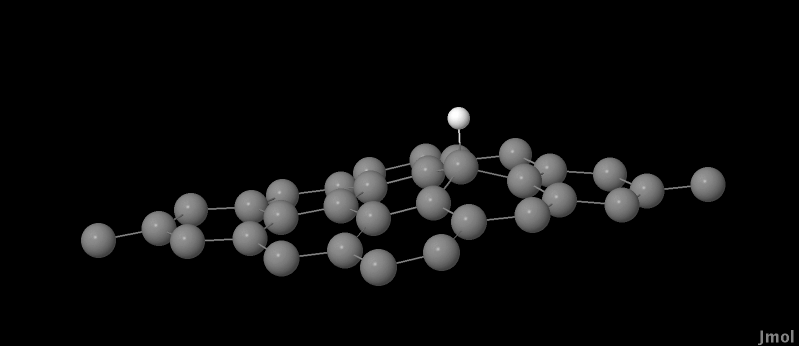}
\caption{
(Color online)
Upper pannel:
Snapshot configuration for a graphene layer subject to
vibrations at $T = 300$ K (an arbitrary time,
$t=260$ fs has been chosen).
Lower pannel:
equilibrium positions on the graphene layer upon
hydrogen adsorption.
}
\label{fgr:GrTyH}
\end{figure}

To quantitatively compare corrugations on graphene layers
originated from other effects, we investigate the effect
of thermal fluctuations ($T \ne 0$ K).
The relationship between stresses originated in free finite edges and thermal 
fluctuations has already been reported in the literature.\cite{huang09}. 
In our simulations, based in periodic models, stresses due to free edges do 
not naturally occur, but it is interesting to notice how our uniform 
compressive strain of $-0.05$ yields results comparable to those simulations. 
To introduce a finite temperature in our simulations we
use ab-initio molecular dynamics to follow the trajectories of
carbon atoms in a  $4 \times 4$ unit cell in equilibrium with a
Nose-Hoover thermostat at $T=300$ K.
Starting out from an equilibrium distribution
the graphene layer is allowed to evolve in the canonical
ensemble for $\tau_{0}=1$ ps using a $\Delta t = 0.5$ fs step to integrate
the equations of motion.
As before, the same two statistical indicators,
$\overline{c}$
and $c_{M}$, are
computed averaging over all the configurations reached inside
the time interval $\tau_{0}$ (the shape of the layer
at a time chosen at random, $t=260$ fs, is shown in the
upper panel of Fig. \ref{fgr:GrTyH}).
Table III shows values for three
strains on the unit cell:
$\epsilon = 0, -0.03$, and $-0.05$.
The last two strains correspond to stresses
of $\sigma_{xx} = \sigma_{yy} = -0.94$ eV/{\AA}$^2$,
and $-1.69$ eV/{\AA}$^2$ respectively.
The case $\epsilon=0$ allows a double check by comparing with
the experimental Debye-Waller factor
determined from low-energy electron diffraction on
graphite (0001):\cite{wu82} 0.053 {\AA} to be compared with
a value of 0.069 {\AA} as derived from the simulation.
Even for the small strain values considered here, it is
clear the non-linear growth of corrugations
vs the external strain.

\subsection{Adsorption of a light impurity.}

A further source for corrugations at the atomic level is
the presence of defects. For the sake of simplicity,
we consider the adsorption of a light element
(hydrogen atom), which is known to originate a
static distortion in the lattice that depends on the coverage, $\theta$,
and extends as a long-range elastic perturbation.\cite{andres08}
For $\theta=\frac{1}{16}$,
a value where the interaction between periodic images of the
adsorbate is already low and can be representative of the
behavior for the isolated impurity, both the averaged and the maximum corrugation
take values fairly similar
to the ones obtained by heating the layer to
$T=300$ K, except perhaps for the value of $c_{M}$, that systematically
takes lower values for the thermal case (Table III).
The corrugations studied here lead to partial $sp^3$ hybridization of the carbon orbitals,
enhancing the spin-orbit coupling.\cite{NG09}

\section{Conclusions.}

We find that averaged atomic corrugations of $\approx 0.05$ {\AA}, and
maximum corrugation values of $\approx 0.5$ {\AA} are easily obtained
under realistic conditions of cleanness and temperature
on graphene layers. Corrugations due to thermal vibrations
of atoms at room temperature
make a similar effect to the adsorption of a light atom (H) with a
moderately small coverage ($\theta=1/16$).
An external compressive stress increases corrugations in a non-linear way.
Due to the flexural mode, the graphene layer
becomes intrinsically unstable and shows a value for the maximum corrugation
under compressive stress
three or four times larger than values attained at $T \le 300$ K or
because the adsorption of impurities.
Upon a 4\% contraction, the LO and TO modes increase their
frequencies by about 20 \%;
while the main elastic constant affected is $c_{12}$,
that is approximately halved.\cite{politano12,cadelano12}
On the other hand, the frequencies of the ZA and ZO modes are 
reduced.
These modes induce an effective $sp^3$ coordination in the lattice, and their softening under strain leads to an enhancement of the intrinsic spin-orbit coupling\cite{NG09}, 
opening a way to observe topological insulator features in graphene.\cite{KM05}
Finally, 
these simulations have been useful to obtain the elastic constants
needed to construct an analytical model of free standing membranes
based in elasticity theory.\cite{AGK12}
Such a simple analytical model yields 
negative Gr\"uneisen parameters related to the bending modes,
and a simple explanation for the negative thermal expansion
coefficient of graphene.

\section{Acknowledgments.}
This work has been financed by the
MICINN, Spain, (MAT2011-26534, FIS2008-00124, FIS2011-23713, CONSOLIDER CSD2007-
00010), and ERC, grant 290846.
MIK acknowledges financial support from FOM, the Netherlands.
Computing resources provided by the CTI-CSIC are
gratefully acknowledged.


%

\clearpage
\newpage

\begin{table}
\label{cijGraphite}
\begin{tabular}{c|lllll}
                   & $c_{11}$      & $c_{12}$         & $c_{13}$   & $c_{33}$    & $c_{44}$  \\ \hline
EXP\cite{cijgrIXR} & $1109 \pm 16$ & $139 \pm 36$ & $0 \pm 3$ & $39 \pm 7$ & $5 \pm 3$  \\
EXP\cite{cijgrPyr} & $1060 \pm 20$ & $180 \pm 20$ & $1.5 \pm 0.5$ & $37 \pm 1$ & $0.3$ \\
REF\cite{SKF11}    & $1109$        & $175$        & $-2.5$        & $29$       & $4.5$ \\
THIS WORK          & $1069$        & $204$        & $-2.8$        & $32$       & $1$ \\
\end{tabular}
\caption{Comparison of ab-initio elastic constants (GPa) computed for graphite with
experimental values determined on single crystals by inelastic x-ray
scattering\cite{cijgrIXR}, on highly oriented pyrolitic graphite\cite{cijgrPyr},
and independent theoretical calculations\cite{SKF11}.
}
\end{table}

\begin{table}
\label{cij2D}
\begin{tabular}{c|cccccc}
$c_{ij}$ & I  & AB   & AA   & AA'  & BULK & EXP \\ \hline
$c_{11}$ & 23 & 22   & 22   & 11   & 22   & 22 \\
$c_{12}$ &  4 &  4   &  4   &  4   &  4   &  4 \\
$c_{66}$ &  9 &  9   &  9   &  3   &  9   &  9 \\
$c_{33}$ &    &      &      &      & 0.25 & 0.28 \\
$c_{44}$ &    &      &      &      & 0.01 & 0.03 \\
$c_{13}$ &    &      &      &      & -0.1 & 0.1  \\
\end{tabular}
\caption{Elastic constants (eV/Ang$^2$) normalized to the carbon
mass in the $1 \times 1$ graphene unit cell.
Results are given for (I) a single graphene
layer, (AB) bilayer with Bernal stacking,
(AA) bilayer with direct stacking,
(AA') meta-estable bilayer at short distances\cite{andres08AA},
(BULK) graphite,
and experimental values for graphite (EXP\cite{ahmadieh73}).
DFT calculations have been performed using LDA and
norm-conserving pseudopotentials as explained in the text.
No attempt to compute elastic constants involving strains in the 
z-direction has been made for super-cells with a vacuum separator.
}
\end{table}

\begin{table}
\label{tbl:phn}
\begin{tabular}{c|cc|cccccc|cccc}
             & G & & M & & & & & & K & & & \\ \hline
Saito et al.\cite{saito} & 49 & 27 & 47 & 43 & 39 & 24 & 20 & 14 & 45 & 39 & 32 & 18 \\
This work                & 48 & 27 & 44 & 41 & 40 & 20 & 19 & 14 & 42 & 37 & 31 & 16 \\
\end{tabular}
\caption{
Comparison of selected phonon frequencies (THz) at high-symmetry
points in the Brillouin zone.
}
\end{table}

\begin{table}
\begin{tabular}{| l c | c c | c c | c c|}
\hline
\multicolumn{2}{|c|}{} &\multicolumn{2}{|c|}{$\epsilon=0.00$} &\multicolumn{2}{|c|}{$\epsilon=-0.03$} &\multicolumn{2}{|c|}{$\epsilon=-0.05$} \\ \hline
CASE&$n \times n$& $\overline{c}$ & $c_{M}$ & $\overline{c}$ & $c_{M}$ &$\overline{c}$ & $c_{M}$ \\\hline
T=300 K         & 4x4         & 0.069 & 0.572 & 0.153 &1.280 & 0.309 & 2.349\\
\hline
H                    & 4x4         & 0.041 & 0.432 &  0.168 & 1.061& 0.310 & 1.580\\
\hline
PHONON ZA & 8x2         & 0.000& 0.000& 0.662  &  2.034 &  0.917 & 2.807  \\
\hline
\end{tabular}
\caption{
Average, $\overline{c}$ ({\AA}),
and maximum, $c_{M}$ ({\AA}), corrugations in a graphene layer
subject to an external thermal bath ($T=300$ K), the
adsorption of a light atom (H), or related to the intrinsic
instability due to its 2D character (mode AZ
in \ref{fgr:PH}).
}
\label{tbl:I}
\end{table}

\end{document}